\documentclass[final,5p,times,twocolumn]{elsarticle}
\usepackage{graphics}

\usepackage{amssymb}
\usepackage{color}
\usepackage{lineno}

\journal{Atroparticle Physics}

\begin{document}

\def\mee{$\langle m_{ee} \rangle$~}
\def\mnu{$\langle m_{\nu} \rangle$~}
\def\mmod{$\| \langle m_{ee} \rangle \|$}
\def\mb{$\langle m_{\beta} \rangle$~}
\def\BBz{$\beta\beta(0\nu)$~}
\def\BBzn{$\beta\beta(0\nu)$}
\def\BBm{$\beta\beta(0\nu,\chi)$~}
\def\BBd{$\beta\beta(2\nu)$~}
\def\BB{$\beta\beta$~}
\def\Mz{$|M_{0\nu}|$~}
\def\Md{$|M_{2\nu}|$~}
\def\Tz{$T^{0\nu}_{1/2}$~}
\def\Td{$T^{2\nu}_{1/2}$~}
\def\Tm{$T^{0\nu\,\chi}_{1/2}$~}
\def\ca{$\sim$}
\def\dca{$\approx$}
\def\dot{$\cdot$}
\def\teod{TeO$_2$~}
\def\thdt{$^{232}$Th }
\def\tldn{$^{208}$Tl}
\def\udt{$^{238}$U~}
\begin{frontmatter}

\title{A novel technique of particle identification with bolometric detectors}

\author[INFNmib]{C.Arnaboldi}
\author[INFNmib,UniMib]{C.Brofferio}
\author[INFNmib]{O.Cremonesi}
\author[INFNmib,UniMib]{L.Gironi\corref{cor1}}
\ead{luca.gironi@mib.infn.it}
\author[INFNmib,UniMib]{M.Pavan}
\author[INFNmib]{G.Pessina}
\author[INFNmib]{S.Pirro}
\author[INFNmib]{E.Previtali}

\cortext[cor1]{Corresponding author: tel: +39 02 64 48 21 07; fax: +39 02 64 48 24 63}

\address[INFNmib]{INFN - Milano Bicocca, Italy}
\address[UniMib]{Dipartimento di Fisica - Universit\`{a} di Milano Bicocca, Italy}

\begin{abstract}

We report in this paper the proofs that the pulse shape analysis can be used in some bolometers to identify the nature of the interacting particle. Indeed, while detailed analyses of the signal time development in purely thermal detectors have not produced so far interesting results, similar analyses on bolometers built with scintillating crystals seem to show that it is possible to distinguish between an electron or $\gamma$-ray and an $\alpha$ particle interaction.
This information can be used to eliminate background events from the recorded data in many rare process studies, especially Neutrinoless Double Beta decay search. Results of pulse shape analysis of signals from a number of bolometers with absorbers of different composition (CaMoO$_4$, ZnMoO$_4$, MgMoO$_4$ and ZnSe) are presented and the pulse shape discrimination capability of such detectors is discussed.

\end{abstract}


\end{frontmatter}

\section{Rare event searches}
\label{RES}


Rare event studies, such as the search for Neutrinoless Double Beta decay (\BBzn) \cite{BBreview} or the identification of Weakly Interacting Massive Particle (WIMP) interactions with ordinary matter \cite{DMreview}, are of extreme interest in astroparticle physics, since they would imply new physics beyond the Standard Model. In both cases, as in all the rare event studies, spurious events are a limiting factor to the reachable sensitivity of the experiment. Unfortunately natural radioactive background is often present in the detector itself or in the materials surrounding it, no matter how much one can try to reduce it with shieldings, selection of materials and complicated purification techniques. In order to handle the residual unavoidable background, all the envisaged approaches require both a good energy resolution (which always helps in the comprehension of the different structures of an energy spectrum) and the capability to identify the nature of the projectile that interacted with the detector. Indeed, the searched event has always well defined signatures helping to distinguish it from background, for instance two electrons with a fixed sum energy in the case of the \BBz. 

Bolometers \cite{bolometers} are based on the detection of phonons produced after an energy release by an interacting particle and can have both an excellent energy resolution and extremely low energy threshold with respect to conventional detectors. They can be fabricated from a wide variety of materials, provided they have a low enough heat capacity at low temperatures, which is the only requirement really unavoidable to build a working bolometer. The latter is a priceless feature for experiments that aim at detectors containing particular atomic or nuclear species to optimize the detection efficiency. If other excitations (such as ionization charge carriers or scintillation photons) are collected in addition to phonons, bolometers have already shown to be able to discriminate nuclear recoils from electron recoils, or $\alpha$ particles from $\beta$ particles and $\gamma$-rays. In this paper we will report on the possibility to obtain similar results just by pulse shape analysis, without the requirement of a double readout for phonons and ionization or scintillation light.

\section{Bolometric Technique and Scintillating Bolometers}
\label{BolTec}

Bolometers can be essentially sketched as a two-component object: an energy absorber in which the energy deposited by a particle is converted into phonons, and a sensor that converts thermal excitations into a readable signal. The absorber must be coupled to a constant temperature bath by means of a weak thermal conductance.

Denoting by C the heat capacity of the bolometer, the temperature variation induced by an energy release E in the absorber can be written as

\begin{equation}
\label{eq:temperature}
\Delta T = \frac{E}{C}
\end{equation}

The accumulated heat flows then to the heat sink through the thermal link and the absorber returns to the base temperature with a time constant $\tau$ = C/G, where G is the thermal conductance of the link:

\begin{equation}
\label{eq:signal}
\Delta T(t) = \frac{E}{C} e^{ - \frac{t}{\tau}}
\end{equation}

In order to obtain a measurable temperature rise the heat capacity of the absorber must be very small: this is the reason why bolometers need to be operated at cryogenic temperatures (of the order of 10-100 mK).

A real bolometer is somewhat more complicated than the naive description presented above. It is made of different elements and it is therefore represented by more than one heat capacity and heat conductance. As such, the time development of the thermal pulse is characterized by various time constants. In principle, if the bolometer performs as an ideal calorimeter and if the conversion of the energy into heat deposited by the particle is instantaneous (as assumed in equation \ref{eq:signal}), then the device is insensitive to the nature of the interacting particle. Although this situation is generally very far from reality, it is however true that the small differences are difficult to detect and the goal has been so far achieved only relying on complicated solutions. Among these are scintillating bolometers.

The concept of a scintillating bolometer is very simple: a bolometer coupled to a light detector \cite{CaF2}. The first must consist of a scintillating absorber thermally linked to a phonon sensor while the latter can be any device able to measure the emitted photons. The driving idea of this hybrid detector is to combine the two available informations, the heat and the scintillation light, to distinguish the nature of the interacting particles, exploiting the different scintillation yield of $\beta$/$\gamma$, $\alpha$ and neutrons.
Dark Matter as well as \BBz searches can benefit of this capability of tagging the different particles, and more generally this technique can be exploited in any research where background suppression or identification is important.

Dark matter experiments look for a very rare signal generally hidden in a huge background. The signal is a nuclear recoil with an energy of few keV (or less) induced by the scattering of a WIMP off a target nucleus. Experiments like CDMS \cite{CDMS}, Edelweiss \cite{Edelweiss} or CRESST \cite{CRESST} clearly show that in such energy region the background is dominated by $\beta$/$\gamma$ interactions. A second source of background are $\alpha$ decays, contributing through energy degraded $\alpha$'s and nuclear recoils. The capability to distinguish a nuclear recoil - candidate for a WIMP interaction - from $\alpha$ or $\beta$/$\gamma$ clearly allows to improve drastically the experimental sensitivity.

A similar approach was proposed also for applications in \BBz searches \cite{CaF2}. More recently, such a possibility has been demonstrated to be viable for a number of candidate nuclei \cite{Pirr06}. In this case the major interest is the identification of $\alpha$ interactions. Indeed the other important source of background, namely $\gamma$-rays, is virtually indistinguishable from the \BBz signal. The suggested way to eliminate the problem of $\gamma$-rays contribution is to study isotopes with a transition energy above 2615~keV. This corresponds in fact to the highest energy $\gamma$-ray line from natural radioactivity and is due to $^{208}$Tl. Above this energy there are only extremely rare high energy $\gamma$'s from $^{214}$Bi (all the \BB active isotopes with Q$_{\beta\beta}>$2615 keV are listed in Tab.~\ref{tab:isotopes}).
Once $\gamma$-rays are no more a worrisome source of background, what is left - on the side of radioactivity- are $\alpha$ emissions. Indeed $\alpha$ surface contaminations not only can represent the dominant background source for \BBz searches based on high transition energy isotopes, but they are already recognized as the most relevant background source in the bolometric experiment CUORICINO~\cite{CUORICINO,CUOpotential,ArtChambery} and as a limiting factor for the experiment CUORE \cite{CUOpotential,CUORE}. Both the experiments search for the \BBz of $^{130}$Te whose transition energy is at 2527 keV, therefore in a region where $\gamma$ background (mainly due to Compton events produced by 2615 keV photons) can be still important.

\begin{table}[]
\begin{center}
\caption{Double beta decay isotopes with endpoint energies above the $^{208}$Tl line.}
\begin{tabular}{ccc}
\hline\noalign{\smallskip}
\space Isotope   \space   & \space Q$_{\beta\beta}$ [MeV] \space & \space natural abundance \space \\
\noalign{\smallskip}\hline\noalign{\smallskip}
$^{116}$Cd & 2.80  & 7.5 \% \\
$^{82}$Se  & 3.00  & 9.2 \%\\
$^{100}$Mo & 3.03  & 9.6 \%\\
$^{96}$Zr  & 3.35  & 2.8 \%\\
$^{150}$Nd & 3.37  & 5.6 \%\\
$^{48}$Ca  & 4.27  & 0.19 \%\\
\noalign{\smallskip}\hline
\end{tabular}
\label{tab:isotopes}
\end{center}
\end{table}

The $\alpha$ contribution to the background in the \BBz region (i.e. at about 3 MeV) is the following. In the natural chains we have various nuclei that decay emitting an $\alpha$ particle with an energy between 4 and 8 MeV, their energy is quite higher than most \BBz Q-values. However, if the radioactive nucleus is located at a depth of a few $\mu$m inside a material facing the detector, the $\alpha$ particle looses a fraction of its energy before reaching the detector and its energy spectrum looks as a continuum between 0 and 4-8 MeV \cite{EPJA}.  A similar mechanism holds in the case of surface contaminations on the bolometer.
 
This radioactive source plays a role in almost all detectors but it turns out to be particularly dangerous for fully active detectors, as in the case of bolometers. It can be efficiently identified and removed with active background suppression technique such as that conceived with scintillating bolometers.

The development of a hybrid detector, able to discriminate $\alpha$ particles and optimized for \BBz searches, was the main purpose of our studies on scintillating bolometers. We tested several devices, differing mainly in the scintillating crystal material and size, to study their thermal response, light yield and radio purity. The results obtained so far are reported in \cite{CDWO4,CAMOO4,ZNSE,ZNMOO4}. 
On our way, we discovered an extremely interesting feature of some of the tested crystals: the different pulse shape of the thermal signals produced by $\alpha$'s and by $\beta$/$\gamma$'s. This feature opens the possibility of realizing a bolometric experiment that can discriminate among different particles, without the need of a light detector coupled to each bolometer. In the case of a huge, multi-detector array, such as CUORE \cite{CUORE} and EURECA \cite{EURECA}, the benefits of employing this technique would be impressive:
 
\begin{itemize}
\item 
more ease during assembly because the single element of the array would be a quite simpler device.
\item
fewer readout channels, with not only an evident reduction of cost and work, but also a cryogenic benefit (in a cryogenic experiment particular care should be devoted to reduce any thermal link between room temperature and the bolometers working at few mK: the heat load of the readout channels must be taken into account and their reduction is always a good solution).
\item
a significant cost reduction, saving money and work that would be necessary for the light detectors procurement and optimization. 
\item
no need of light collectors, this would simplify the structure of the assembly and it would allow the use of coincidences between facing crystals to further reduce the background.
\end{itemize}

As a final remark, it is worth to be mentioned that these devices could be used also for the measurement of $\alpha$ emissions from surfaces, when extremely low counting rates are needed. Indeed, due to their lack of a dead layer and their high energy resolution, bolometers have an extraordinary sensitivity to low range particles like $\alpha$'s. However, a conventional bolometer cannot distinguish the nature of the interacting particle. It provides therefore only a limited diagnostic power (especially for $\alpha$ particles with energies lower than 2615~keV where the $\beta$/$\gamma$ induced background dominates the detector counting rate). On the other hand, a scintillating bolometer has to be surrounded by a reflector to properly collect the scintillation light (therefore cannot be faced to a sample whose radioactive emission has to be identified). The devices here discussed overcome these two difficulties. Traditionally the devices used in this field are Si surface barrier detectors. For low counting rates, large area low background detectors are needed. Today Si surface barriers detectors with an active area of about 10 cm$^2$, a typical energy resolution of about 25-30 keV FWHM, and counting rates of the order 0.05 count/h/cm$^2$ between 3 and 8 MeV are available \cite{CANBERRA}. A bolometer like those here discussed can easily reach a much larger active area, has a typical energy resolution of 10 keV and a background counting rate in the 3-8 MeV region that can be as low as 0.001 count/h/cm$^2$. Thanks to the particle identification technique discussed in this paper, it can distinguish an $\alpha$ emission from a  $\beta/\gamma$ one\footnote{This feature is extremely important when the sample which is investigated produces a continuous counting rate which could be due either to $\alpha$ emissions from a thick contamination or to a $\beta/\gamma$ continuum.} and finally can reject the $\beta$/$\gamma$ background extending its measurement field to energies by far lower than 3 MeV. 

In the following section we report the results obtained with the pulse shape analysis on some of the tested crystals.

\section{Detectors, set-up and data analysis}

The results discussed in this paper have been obtained operating different scintillating bolometers in an Oxford 200 $^{3}$He/$^{4}$He dilution refrigerator located deep underground, in the National Laboratory of Gran Sasso (L'Aquila, Italy). The rock overburden (average depth $\sim$3650 m.w.e. \cite{Hime}) ensures a strong suppression of cosmic rays that in our case is mandatory to be able to operate the detectors without an overwhelming pile-up. A detailed description of the experimental setup can be found in \cite{SETUP}. In order to study the pulse shape characteristics of different materials (in particular those of interest for \BBz), we operated a number of  scintillating bolometers differing for size, geometry and, of course, the absorbing material. As light detector we have used a second bolometer able to absorb scintillation photons converting their energy into heat. This was realised using as absorber Ge wafers of about 5 g, covered - on the side facing the scintillating crystal -  with a 600~\AA\ thick layer of SiO$_2$ in order to increase the light absorption. In this way they provided measurable thermal signals over an extremely large band of scintillation wavelengths.

Both the scintillating crystal and the Ge wafer were equipped with a Neutron Transmutation Doped Ge thermistor (NTD) \cite{NTD}, glued on the crystal surface and used as a thermometer to measure the heat or light signal produced by particles traversing the scintillating crystal.

A silicon resistance, glued on the crystals, was used to produce a calibrated heat pulse in order to monitor the thermal gain of the bolometer. This is indeed subject to variation upon temperature drifts of the cryostat that can spoil the energy resolution. In most cases this temperature drift could be re-corrected off-line on the basis of the measured thermal gain variation \cite{ALES98,ARNA03}.
  
\subsection{Read-out and DAQ}

The read-out \cite{ELE} of the thermistor was performed via a preamplifier stage, a second stage of amplification and an antialiasing filter (a 6 pole roll-off active Bessel filter 120 db/decade \cite{Bessel}) located in a small Faraday cage. The ADC was a NI USB-6225 device (16 bit 40 differential input channels). For each triggered signal the entire waveform (\emph{raw-pulse}) is sampled, digitized and acquired for the off-line analysis. 
Since all the relevant parameters (including the amplitude) of the triggered signals are evaluated off-line, a particular care has to be dedicated to the optimization of the signal filtering and digitization.
In the case of the scintillating bolometer the large heat capacity of the absorber, coupled to the finite conductance of the crystal-glue-thermistor interface results in quite slow signals, characterised by a rise-time of the order of few ms\footnote{The minimum observable signal rise-time is limited by the integration on the parasitic capacitance of the signal wires that connect the NTD thermistor to the front-end electronics.} and a decay-time of hundreds of ms (determined by the crystal heat capacity and by its thermal conductance toward the heat sink). 
Consequently the sampling rate typically used for the signal is 1-4 kHz, over a time window of 200-2000~ms. The Bessel filter acts mainly as antialising, to avoid spurious contributions in the sampled signal. Generally it is preferred to fix its cut-off frequency at the lowest value that does not deteriorate the signal to noise ratio (i.e. to obtain the best results in terms of energy resolution). This results to be a frequency of the order of 10~Hz, which is by far lower than what needed for antialiasing purposes. In the studies here presented the Bessel cut-off frequency was fixed at 120~Hz in order to exploit the maximum available information in the signal bandwidth.

\subsection{Analysis techniques}
\label{AnaTech}

Off-line analysis aims at determining the pulse amplitude and energy together with several pulse shape parameters associated with each raw-pulse waveform recorded by the data acquisition system. Starting from these quantities the physical informations that are relevant for the scientific goals can be extracted.

The first step of the analysis consists in the correct evaluation of the pulse amplitude. Since thermal pulses are superimposed to stochastic noise, a simple maximum-minimum algorithm would not give the better achievable resolution. We therefore use the Optimum Filter (OF) approach \cite{GATTI86}. This algorithm has proven to provide the best estimate of the pulse amplitude and, as a consequence, the best energy resolution. The basic concept is to build a filter that, when applied to the raw-pulse, produces - as output - a pulse  with the best signal to noise ratio. The filtered pulse is then used to evaluate the signal amplitude. It can be proven that in the frequency domain the OF transfer function H( $\omega$ ) is given by 

\begin{equation}
\label{eq:OF}
H(\omega) = K ~ \frac{S^*(\omega)}{N(\omega)} ~ e^{-j\omega t_M} 
\end{equation}  

\noindent where S($\omega$) is the Fourier transform of the ideal thermal signal (reference pulse in the absence of noise), N($\omega$) is the noise power spectrum, $t_M$ is the delay of the current pulse with respect to the reference pulse and K is a proper normalising factor usually chosen in order to obtain the correct event energy. 

The role of the optimum filter is to weight the frequency components of the signal in order to suppress those frequencies that are more affected by noise. It can be seen from eq. \ref{eq:OF} that, in order to build the filter, the shape of the reference pulse S($\omega$) and the noise power spectrum N($\omega$) must be known. S($\omega$) is usually estimated by averaging a large number of recorded raw-pulses, so that the noise associated with each of them averages to zero. N($\omega$) is obtained according to the Wiener-Khintchine theorem by acquiring many detector baselines in absence of thermal pulses and averaging the corresponding noise power spectra. 

Once the pulse amplitude has been evaluated, gain instability corrections are applied to data. Due to the dependence of the detector response on the working temperature, the same amount of released energy can produce thermal pulses of different amplitudes. Gain instabilities are corrected monitoring the time behaviour of thermal pulses of fixed energy, generated every few minutes across a Si heater resistor attached on the crystal absorber \cite{ALES98,ARNA03}. Finally the amplitude to energy conversion (calibration) is determined by measuring the pulse amplitudes corresponding to fixed calibration lines. In the measurements here reported the signal of the scintillating bolometer (we will refer to this signal as to the heat or thermal signal) has been calibrated on the basis of the full energy peaks visible in the spectrum collected when the detector was exposed to an (external to the cryostat) \thdt source. These peaks have a nominal energy of: 511, 583, 911, 968 and 2615 keV. Below 511~keV and above 2615~keV the energy calibration is extrapolated. However, the heat response for $\alpha$ particles is slightly different from the $\beta$/$\gamma$ response in scintillating bolometers \cite{CDWO4}. For the molybdates that are here reported this heat quenching factor is lower than few percent while for ZnSe it is a little bit higher, about 10 percent for $\sim$6 MeV alpha particles \cite{ZNSE}. This miscalibration of the $\alpha$ band however does not imply an appreciable change in the discrimination confidence level described below.

In the case of the light signal the energy calibration is  not needed and we therefore present its value in arbitrary units. 

Besides the amplitude, few other characteristic parameters of the pulse are computed by the off-line analysis. Some of them are: $\tau_{rise}$ and $\tau_{decay}$, TVL and TVR. The rise-time ($\tau_{rise}$) and the decay-time ($\tau_{decay}$) are determined on the recorded raw-pulse as (t$_{90\%}$-t$_{10\%}$) and (t$_{30\%}$-t$_{90\%}$) respectively. TVR (Test Value Right) and TVL (Test Value Left) are computed on the optimally filtered pulse A(t). They are the root mean square differences between the current signal A(t) and the reference pulse after OF filtering A$_{0}$(t)= H(t)~$\otimes$~S(t). In more detail, the filtered response function A$_{0}$(t) is synchronized with the filtered signal A(t), making their maxima to coincide, then the least square differences of the two functions are evaluated on the right (TVR) and left (TVL) side of the maximum on a proper time interval which is usually chosen  depending on the shape of the OF signals. Although these two parameters do not have a direct physical meaning, however they are very sensitive (even in noisy conditions) to any difference between the shape of the analyzed pulse and the response function. Consequently, they are used either to reject fake triggered signals (e.g. spikes) or to identify variations in the pulse shape with respect to the reference response function (and this will be our case).

\begin{figure*}[ht]
\begin{center}
\includegraphics[width=1.\linewidth]{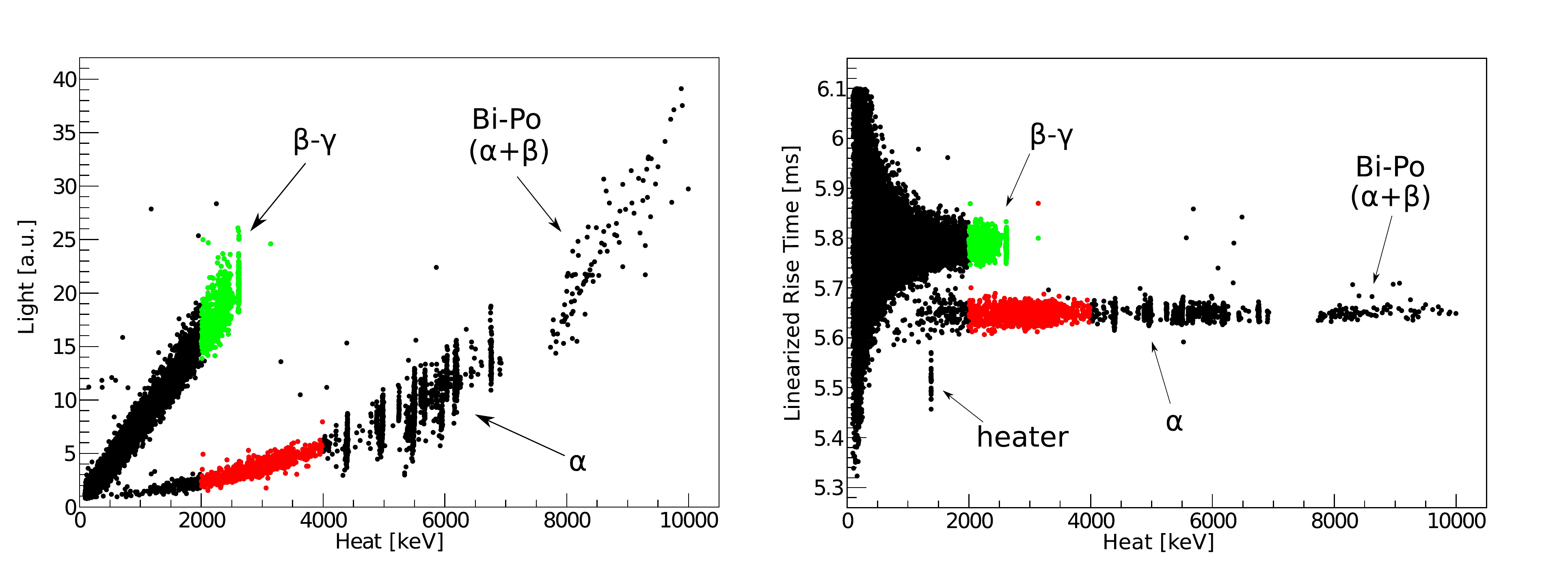}
\end{center}
\caption{CaMoO$_4$ crystal faced to two $\alpha$ sources. Left panel: scatter plot of Light vs. Heat. The energy calibration is performed on the $\gamma$ peaks. In red (light in the bottom curve) events in the 2-4 MeV region due to $\alpha$ particles while in green (light in the top distribution) events due to $\beta$/$\gamma$ in the same region. Right panel: Scatter plot of $\tau_{rise}^{lin}$ vs. Heat in CaMoO$_{4}$ crystal obtained for the same events. In green (light in the upper curve) events in the 2-4 MeV region due to $\beta$/$\gamma$ particles and in red $\alpha$ events.}
\label{fig:camoo4}
\end{figure*}

\section{Pulse shape signature in the heat pulse}
\label{PSA}

A series of measurements was carried out, in which different scintillating bolometers, each coupled to a light detector (but in one case), were exposed to $\gamma$ and $\alpha$ sources. This allowed us to study the response of our devices to different radiations. The light signal was used to identify - on the basis of the heat to light ratio - the particle producing the event under study.

As mentioned in section \ref{AnaTech}, for each triggered signal different pulse shape parameters are computed by the off-line analysis, generally to isolate spurious and pile-up events. In the case of scintillating bolometers, looking at the distribution of  the pulse shape parameters for the heat signals, we realized that it was possible to distinguish $\beta$/$\gamma$ from $\alpha$ events. This is clearly evident in Fig.~\ref{fig:camoo4} where the scatter plot of the amplitudes measured for the light and heat signals (Light vs. Heat) is compared with the scatter plot (obtained for the same events) of the linearized rise-time (see later in the text) vs. amplitude for the heat signal ($\tau_{rise}^{lin}$ vs. Heat).
In this detector, $\alpha$ and $\beta$/$\gamma$ interactions draw different distributions in both the scatter plots, definitely proving that the shape of the thermal pulse induced by an $\alpha$ particle is different from that of a $\beta$/$\gamma$ interaction. 

This behavior can be explained by the dependence of light yield on the nature of the interacting particle. The high ionization density of $\alpha$ particles implies that all the scintillation states along their path are occupied. This saturation effect does not occur or at least is much less for $\beta$/$\gamma$ particles. Therefore, in $\alpha$ interactions a larger fraction of energy flow in the heat channel with respect to $\beta$/$\gamma$ events. This leads not only to a different light and heat yield but also to a different time evolution of both signals. The pulse shape of the thermal signal then can be explained by the partition of energy in the two channels with different decay constants. In particular, as shown by \cite{LightDep1,LightDep2} the scintillation produced in molybdates by $\alpha$ and $\beta$/$\gamma$ particles presents few decay-time constants with different relative intensities. Also a strong temperature dependence of the averaged decay-time of the light pulses in these same crystals was reported \cite{TimeDep1,TimeDep2}. For this reason - while at room temperature the averaged decay-time of molybdate or tungstate scintillators is of the order of tens of $\mu$s (thus almost instantaneous in the heat pulse timescale) - at low temperatures it increases to hundreds of $\mu$s and is therefore comparable to the typical rise-time of the heat signal of our scintillating bolometers.
 
In the following sections we analyse the results obtained with different crystals, each being a possible candidate for a \BBz experiment. We quote for each crystal the discrimination power achieved - in the \BBz region - between $\alpha$ and $\beta$/$\gamma$ particles on the basis of the light/heat ratio or simply on the basis of the pulse shape of the heat signal. We discuss in more detail the case of CaMoO$_4$, briefly summarizing the results obtained for other crystals.

\begin{figure}[ht]
\begin{center}
\includegraphics[ width=1.\linewidth]{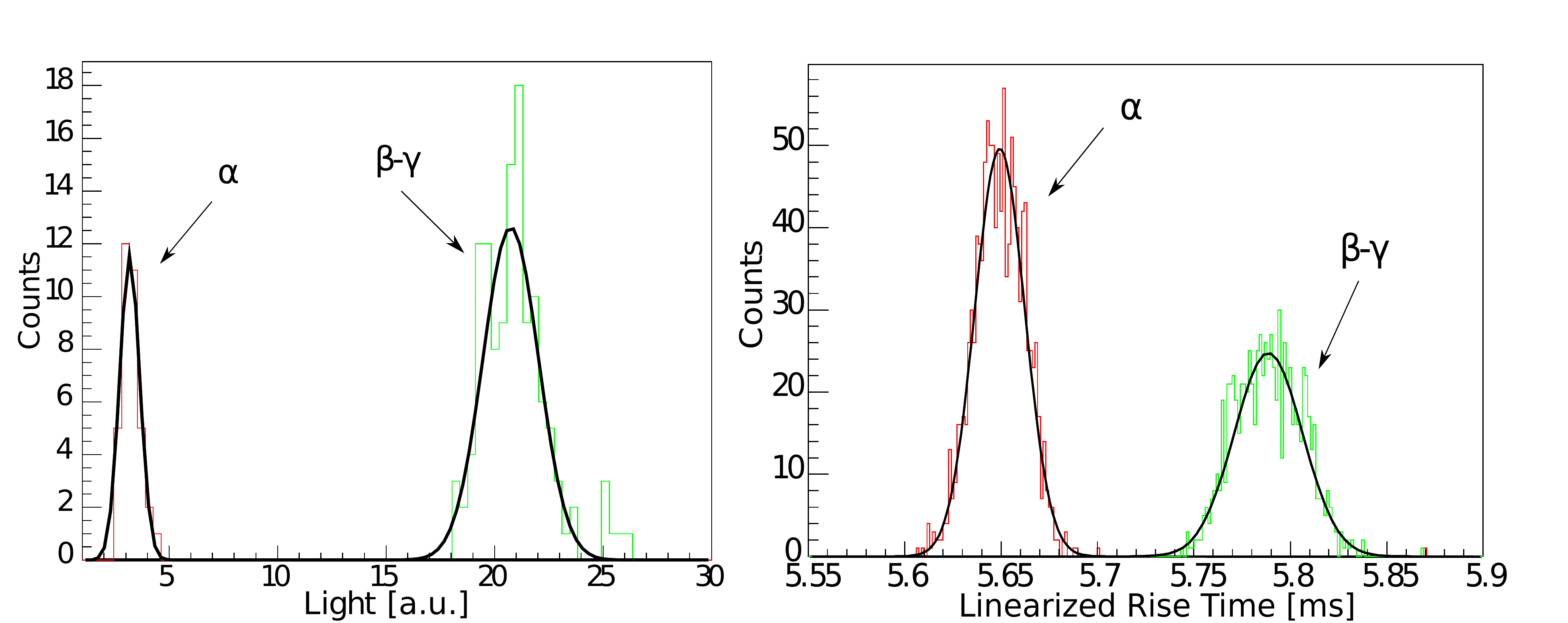}
\end{center}
\caption{Evaluation of the $\beta/\gamma$ and $\alpha$ discrimination power in CaMoO$_4$. Left panel: projection on the Light axis of the $^{208}$Tl $\gamma$-line (2615 keV) (in green on the right) and of $\alpha$ events (in red on the left) that released a similar energy in the CaMoO$_{4}$ crystal. The separation corresponds to 12.6 sigma. Right panel: projection on the $\tau_{rise}^{lin}$ axis of events in the 2-4 MeV regions due to  $\beta$/$\gamma$ events (in green on the right) and $\alpha$ events (in red on the left). The separation here corresponds to 6.5 sigma. These same events are reported, with identical colors, in Fig.~\ref{fig:camoo4}.}
\label{fig:camoo4_light}
\end{figure}

\section{CaMoO$_{4}$}
\label{CaMoO4}

Recently CaMoO$_{4}$ has been intensively studied, for its possible application as a scintillating bolometer for \BBz and Dark Matter experiments \cite{Pirr06,Seny06,Mikh06-JPDAP}. This crystal contains two isotopes that could undergo \BBz: $^{48}$Ca (Q$_{\beta\beta}$=4.27 MeV) and $^{100}$Mo (Q$_{\beta\beta}$=3.03 MeV). Actually, while the large content of $^{100}$Mo makes this crystal very attractive, the presence of $^{48}$Ca is a problem. Indeed, the natural isotopic abundance of $^{48}$Ca (a.i.=0.19\%) is too low to study the \BBz without enrichment, which is extremely difficult, and at the same time, it is too high to study the \BBz of $^{100}$Mo, since the background due to the \BBd of $^{48}$Ca in the \BBz region of $^{100}$Mo will limit the reachable sensitivity for the latter isotope. A possible solution to this problem was suggested by Annenkov et al. \cite{CAMOO4_DEP} who proposed an experiment with CaMoO$_{4}$ depleted in $^{48}$Ca. Despite these possible problems, we discovered that CaMoO$_{4}$ is an extremely interesting crystal because of its capability to discriminate $\beta$/$\gamma$ from $\alpha$, thanks to the different shape of the thermal pulses. 

The sample we used was a cylindric CaMoO$_{4}$ crystal with a mass of 158~g (h = 40mm, $\varnothing$ = 35mm). The crystal was faced to two $\alpha$ sources. Source A was obtained by implantation of $^{224}$Ra in an Al reflecting stripe. The shallow implantation depth allows to reduce to a minimum the energy released by the $\alpha$'s in the Al substrate so that monochromatic $\alpha$ lines (those produced in the decay chain of $^{224}$Ra to the stable $^{208}$Pb isotope) can be observed in the scintillating bolometer. Besides these $\alpha$ particles - all with energies above 5 MeV - the source emits a $\beta$ with a maximum energy of 5 MeV, due to the decay of \tldn.
Since our main goal was to study the efficiency of $\alpha$ particle rejection in the \BBz region (i.e. at about 3 MeV), a second source (B) was also used. This was obtained contaminating an Al stripe with a \udt liquid solution and later covering the stripe with an alluminated Mylar film (6~$\mu$m thick). Thus the source produced a continuous spectrum of $\alpha$ particles, extending from about 3 MeV down to 0. 

In Fig.~\ref{fig:camoo4} we show the Light vs. Heat scatter plot, collected while the crystal was exposed to an external \thdt source. The total live time of this measurement was $\sim$43 h. The FWHM energy resolution, on the heat signal, ranges from 2.7 keV at 243 keV to 8.7 keV at 2615 keV on $\beta$/$\gamma$ events and is about 10 keV on 5 MeV $\alpha$'s. The light yield of the crystal, evaluated calibrating the light detector with a $^{55}$Fe source, is 1.87 keV/MeV (for more details on the procedure used for the evaluation of the light yield see references \cite{CDWO4,ZNSE}).

The two separate bands - clearly visible in the Light vs. Heat scatter plot - are ascribed to $\beta$/$\gamma$'s (upper band) and $\alpha$'s (lower band). The upper band is dominated by the \thdt source  $\gamma$'s (plus the environmental $\gamma$'s). The lower band is due to $\alpha$'s from source A and from Uranium and Thorium internal contamination of the crystal (these are the monochromatic lines above 4 MeV) and source B (the continuum counting rate below 4 MeV). Above 8~MeV we observe a group of events ascribed to $\alpha$+$\beta$ summed signals due to internal contamination in Bismuth and Polonium. Indeed, due to the long rise-time of this device (\ca5 ms), the beta decay of $^{214}$Bi or $^{212}$Bi (respectively of $^{238}$U and $^{232}$Th chains) followed immediately by $\alpha$ decay of $^{214}$Po ($\tau$=164 $\mu$s) and $^{212}$Po ($\tau$=298 $\mu$s ), may lead to a pile up on the rise-time of the thermal pulses that can hardly be recognized as a double signal. The two decays produce therefore a single pulse, with an energy that is the sum of the two.

The discrimination between the $\alpha$ and the $\beta$/$\gamma$ populations, provided by the scintillation signal, can be evaluated by measuring the difference between the average amplitude of the light signal produced by the two kinds of particles  (Light$_{\beta/\gamma}$ and Light$_{\alpha}$), considering a group of events releasing the same energy in the scintillating crystal. This difference is then compared with the width of the two distributions ($\sigma_{\beta/\gamma}$ and $\sigma_{\alpha}$). The discrimination confidence level D$_{Light}$ can be then defined as: 
\begin{equation} 
D_{Light} = \frac{Light_{\beta/\gamma}-Light_{\alpha}}{\sqrt{\sigma_{\beta/\gamma}^2+\sigma_{\alpha}^2}}
\end{equation} 
To evaluate this discrimination power in the \BBz region we selected events belonging to the 2615 keV $\gamma$-line full energy peak and compared their light pulse distribution with that of events of similar energy in the $\alpha$ band (Fig. \ref{fig:camoo4_light} left panel). D$_{light}$ results to be 12.6 sigma. 
 The reason for using events with the same energy to evaluate D$_{light}$ is in the energy dependence of the distance (and width) of the $\alpha$ and $\beta$/$\gamma$ bands, that induce an energy dependence of the discrimination confidence level. In the energy range where both the bands are populated (i.e. between 1 and 3 MeV), D$_{Light}$(E) appears to be linearly decreasing with energy. Its extrapolated value at E=0 being 3 sigma.
 
\begin{figure}[th]
\begin{center}
\includegraphics[ width=1.\linewidth]{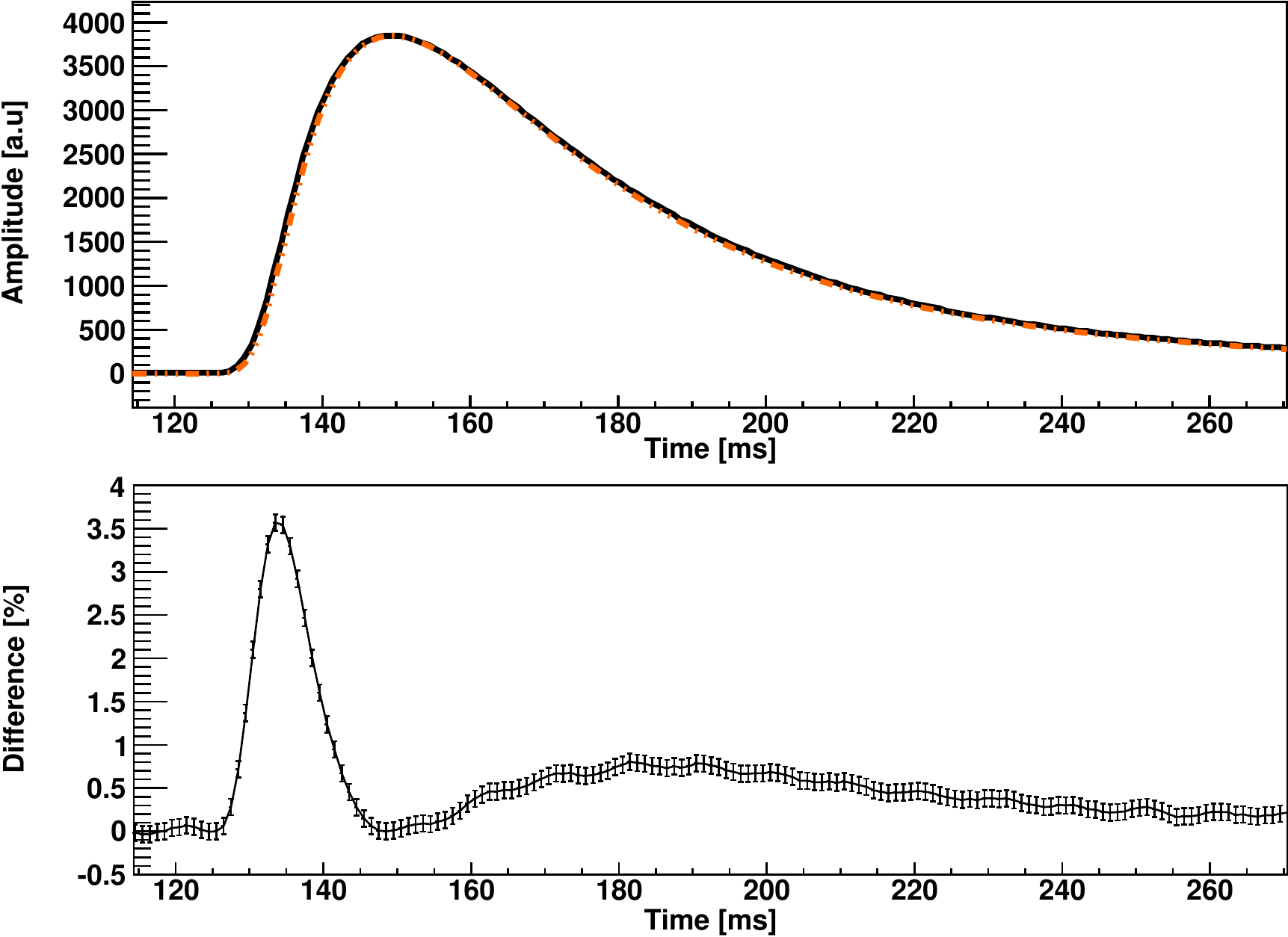}
\end{center}
\caption{CaMoO$_4$ crystal, comparison of $\beta$/$\gamma$'s and $\alpha$'s pulse shape. Top panel: thermal pulse obtained by averaging signals belonging to the 2615 keV $\gamma$-line (black continuous line) and signals due to $\alpha$ particles that release a similar energy (red dotted line). Bottom panel: difference, in percent of the maximum amplitude, between the two averaged pulses.}
\label{fig:Impulso}
\end{figure}

As already anticipated, we discovered that the heat signal shape is enough to discriminate $\beta$/$\gamma$'s from $\alpha$'s. This is shown in the right panel of Fig.~\ref{fig:camoo4} where we report the (thermal pulse) $\tau_{rise}^{lin}$ vs. Heat scatter plot for the same events shown in the left panel. Two separate bands, ascribed to $\beta$/$\gamma$ and $\alpha$ events can be identified. The former with an average rise-time of \ca5.8~ms, the latter with \ca5.6~ms.
In this plot the weak energy (Heat) dependence of $\tau_{rise}$ observed for the two populations was corrected by fitting their rise-time distributions with two lines having the same slope. To do this we used  $\beta$/$\gamma$ events in the 0.5-2.6 MeV range and $\alpha$ events in the 1.5-7 MeV range. The $\tau_{rise}$ is then linearized re-defining it as $\tau_{rise}^{lin}=\tau_{rise} - slope \times E$ with $slope$=0.0075~ms/MeV.

In order to emphasize the correspondence in the identification of $\beta$/$\gamma$ and $\alpha$ events, signals selected in the Light vs. Heat scatter plot as having an energy between 2 and 4 MeV and belonging to the $\beta$/$\gamma$ or $\alpha$ bands, are reported in different colors. We will refer in the following to these two groups of events as to 2-4~MeV $\gamma$'s and 2-4~MeV $\alpha$'s, although - above the 2615~keV line - the $\gamma$ group is empty. These two samples will be used to evaluate the difference in shape among $\beta$/$\gamma$ and $\alpha$ signals. 

\begin{figure*}
\begin{center}
\includegraphics[ width=1.\linewidth]{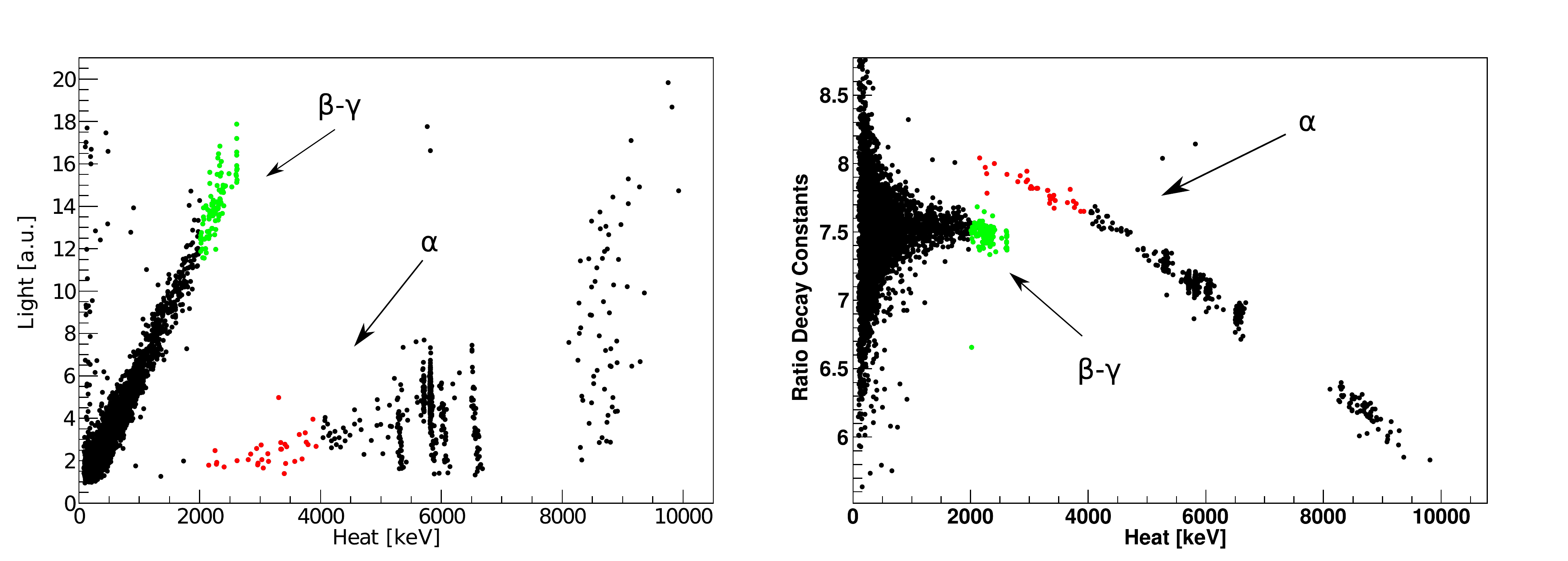}
\end{center}
\caption{ZnMoO$_{4}$ crystal. Scatter plot of Light vs. Heat (left panel) and scatter plot of RDC vs. Heat (right panel). In green events in the 2-4 MeV regions due to $\beta$/$\gamma$ particles and in red those due to $\alpha$'s.}
\label{fig:znmoo4}
\end{figure*}

In Fig.~\ref{fig:Impulso} we compare two pulses obtained by averaging - separately - signals belonging to the 2615 keV $\gamma$-line and signals due to $\alpha$ particles that release a similar energy. The average is here needed to get rid of the noise that can mask the small differences among the pulses. The difference in shape can be appreciated in the bottom panel of Fig.~\ref{fig:Impulso}.

The discriminating power D$_{RiseTime}$ of the heat pulse shape method can be defined exactly as done for D$_{Light}$:
\begin{equation} 
D_{RiseTime}(E) = \frac{RiseTime_{\beta/\gamma}-RiseTime_{\alpha}}{\sqrt{\sigma_{\beta/\gamma}^2+\sigma_{\alpha}^2}}
\end{equation} 

\noindent where, however, the dependence on energy is here due only to the variation of the width of the rise-time distributions since the difference in rise-time appears to be independent on energy. The distribution obtained projecting the rise-time of 2-4 MeV $\gamma$'s and 2-4 MeV $\alpha$'s is shown in Fig. \ref{fig:camoo4_light} (right panel). A gaussian fit of the two peaks yields a rise-time of (5.788$\pm$0.017)~ms for $\gamma$'s and of (5.649$\pm$0.013)~ms for $\alpha$'s. D$_{RiseTime}$ results equal to 6.5 sigma. As for the D$_{Light}$ case, we can extrapolate the D$_{RiseTime}$ value also in the energy region where the $\alpha$ band is poorly populated, simply assuming that the distance between the two bands remains constant (as we observe) and  D$_{RiseTime}$ changes because the width of the two distributions becomes larger, at low energies, due to noise. The result is that D$_{RiseTime}$ becomes lower than 2 sigma at 500 keV. In other words the \ca150 $\mu$s difference in rise-time of $\alpha$ and  $\beta$/$\gamma$ events cannot be appreciated when the pulse amplitude is too low and, therefore, the noise modifies appreciably the signal shape. 
Finally, although in this measurement the rise-time is the most efficient shape parameter for $\alpha$ event discrimination, good discrimination levels were observed also in other parameters such as the decay-time and the TVR. The possibility to increase the discrimination power  by combining the information from different shape parameters or the fit on the rise-time is under study.

We note that either in the Light vs. Heat scatter plot as in the $\tau_{rise}^{lin}$ vs. Heat one we can see some outliers that could indicate a possible failure of the particle identification technique. However such events can be accounted for if one considers the following two effects. If a $\gamma$-ray interacts both in the light detector and in the scintillator the light signal that is read-out has a wrong amplitude since it is not only ascribed to scintillating photons but also to a direct $\gamma$ interaction in the Ge wafer. The measurement is performed in a high rate condition, therefore we expect to have a number of pile-up events in each of the two detectors, this leads to an erroneous evaluation of the pulse amplitude and pulse rise-time.
 
To conclude, in the case of \BBz decay the discrimination power provided by the use of the scintillation signal is comparable with that provided by the pulse shape analysis. For what concerns the use of this device for the measurements of $\alpha$ particle emissions from an external sample, a D$_{RiseTime}$ better than 2 sigma above 500 keV means that this detector has a good sensitivity even in the region where Si surface barrier detectors start to be dominated by $\gamma$ background. On the contrary the applicability of this technique to Dark Matter searches cannot be proved directly.

\subsection{Other crystals}

Other scintillating crystals have shown the possibility to discriminate interacting particles through the thermal pulse shape differences. Among the tested crystals, other molybdates (ZnMoO$_{4}$, MgMoO$_{4}$) and also other crystals such as ZnSe showed a good discrimination power. 

\subsection{ZnMoO$_{4}$}

We tested a 19.8~g ZnMoO$_{4}$ crystal, having the shape of a prism with height of 11mm and a regular hexagonal base, with a diagonal of 25~mm \cite{ZNMOO4}. The FWHM energy resolution, on the heat signal, is 4.2 keV FWHM on $\beta$/$\gamma$ events at 2615~keV, and about 6 keV on the 5.4 MeV $\alpha$ line. The light yield of the crystal is 1.1 keV/MeV. The total live time of this measurement was $\sim$195 h. Also for this analysis we have used the Light vs. Heat scatter plot to select $\beta$/$\gamma$ and $\alpha$ between 2 and 4 MeV, in order to evaluate the discrimination power provided by pulse shape studies.

Unlike for the CaMoO$_4$ case where the most efficient parameter for the discrimination is the rise-time, this measurement showed a higher discrimination power on the decay-time of the thermal pulse. In order to emphasize the discrimination power, all signals have been fitted with a function obtained as the sum of two exponentials:

\begin{equation}
\label{eq:vs. energy}
\Delta V(t) = (e^{ - t/ \tau_1 + A_1} + e^{ - t/ \tau_2 + A_2})
\end{equation}

\noindent where the $\tau_1$ and $\tau_2$ parameters are obtained by fitting the decay of the thermal signals in the raw-pulse acquired for each event. It was observed that the best discrimination power is obtained by the ratio of the two decay constants (RDC):

\begin{equation}
\label{RDC_def}
RDC = \frac{\tau_1}{\tau_2}
\end{equation}

\begin{figure}
\begin{center}
\includegraphics[ width=1.\linewidth]{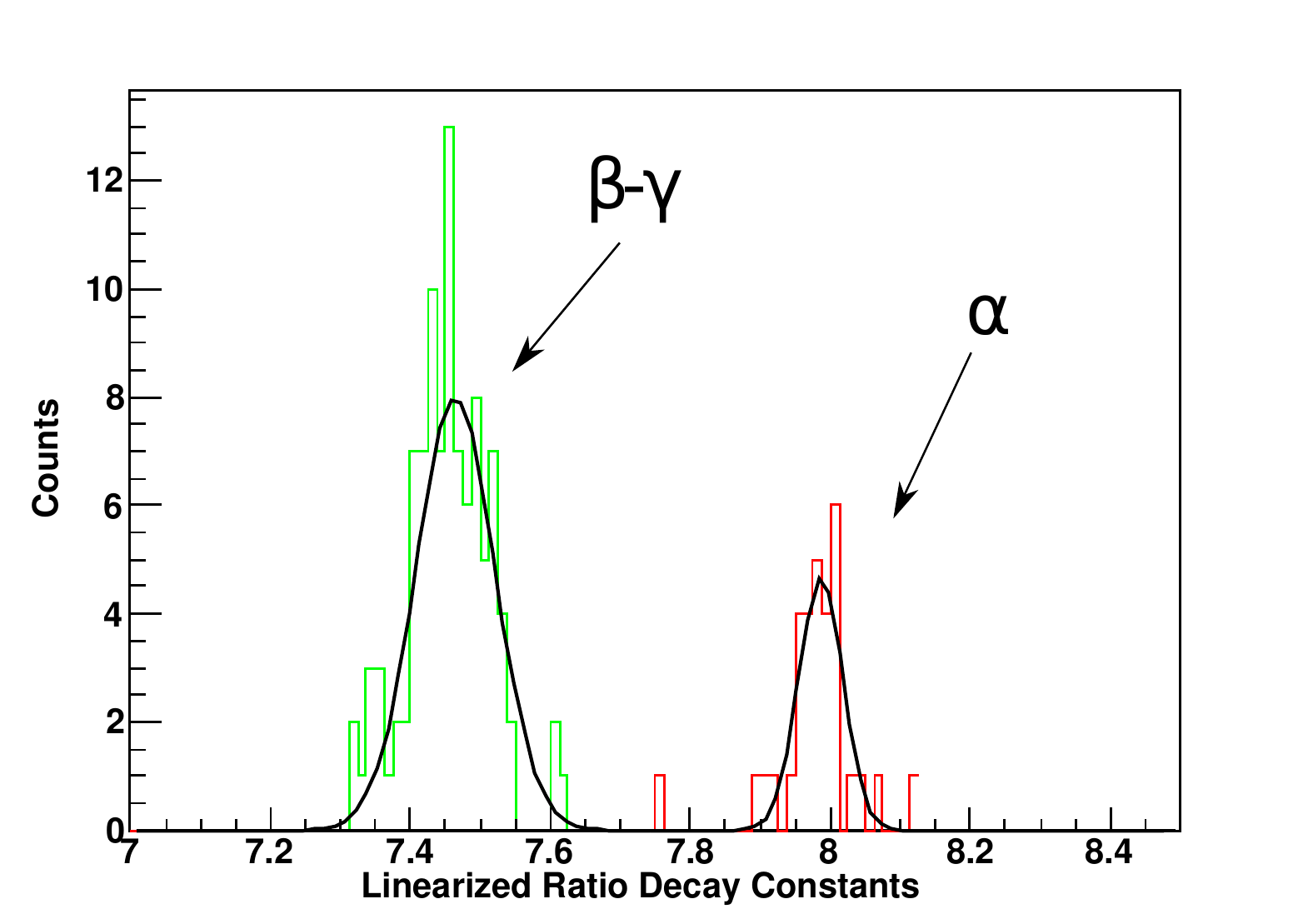}
\end{center}
\caption{Evaluation of the $\beta/\gamma$ and $\alpha$ discrimination power in ZnMoO$_4$. Projection on the RDC$^{lin}$ axis (after linearization of the scatter plot of Fig. \ref{fig:znmoo4}, right panel) of events in the 2-4 MeV region due to $\beta$/$\gamma$ events (in green) and $\alpha$ events (in red).}
\label{fig:ZnMoO4_RDC_Proj}
\end{figure}

\noindent The scatter plot of RDC vs. Heat is reported in the right panel of Fig. \ref{fig:znmoo4} (in the left panel the corresponding Light vs. Heat scatter plot). The discrimination between $\alpha$ and $\beta$/$\gamma$ populations provided by RDC is evaluated linearizing the RDC vs. Heat relationship (fitting the $\alpha$ band in scatter plot of Fig. \ref{fig:znmoo4} with a polynomial as done for the $\tau_{rise}$ of CaMoO$_4$) and then projecting the distribution of RDC$^{lin}$ for events in the 2-4 MeV range. The two peaks (${\beta/\gamma}$ and ${\alpha}$) are then fitted with a gaussian (Fig. \ref{fig:ZnMoO4_RDC_Proj}). The discrimination power D$_{RDC}$ is 6.4 sigma.

\subsection{MgMoO$_{4}$}

This compound contains, as in the two previous cases, the \BB active isotope $^{100}$Mo which is here present in a larger concentration  (\ca52\% in mass).

\begin{figure}
\begin{center}
\includegraphics[ width=1.\linewidth]{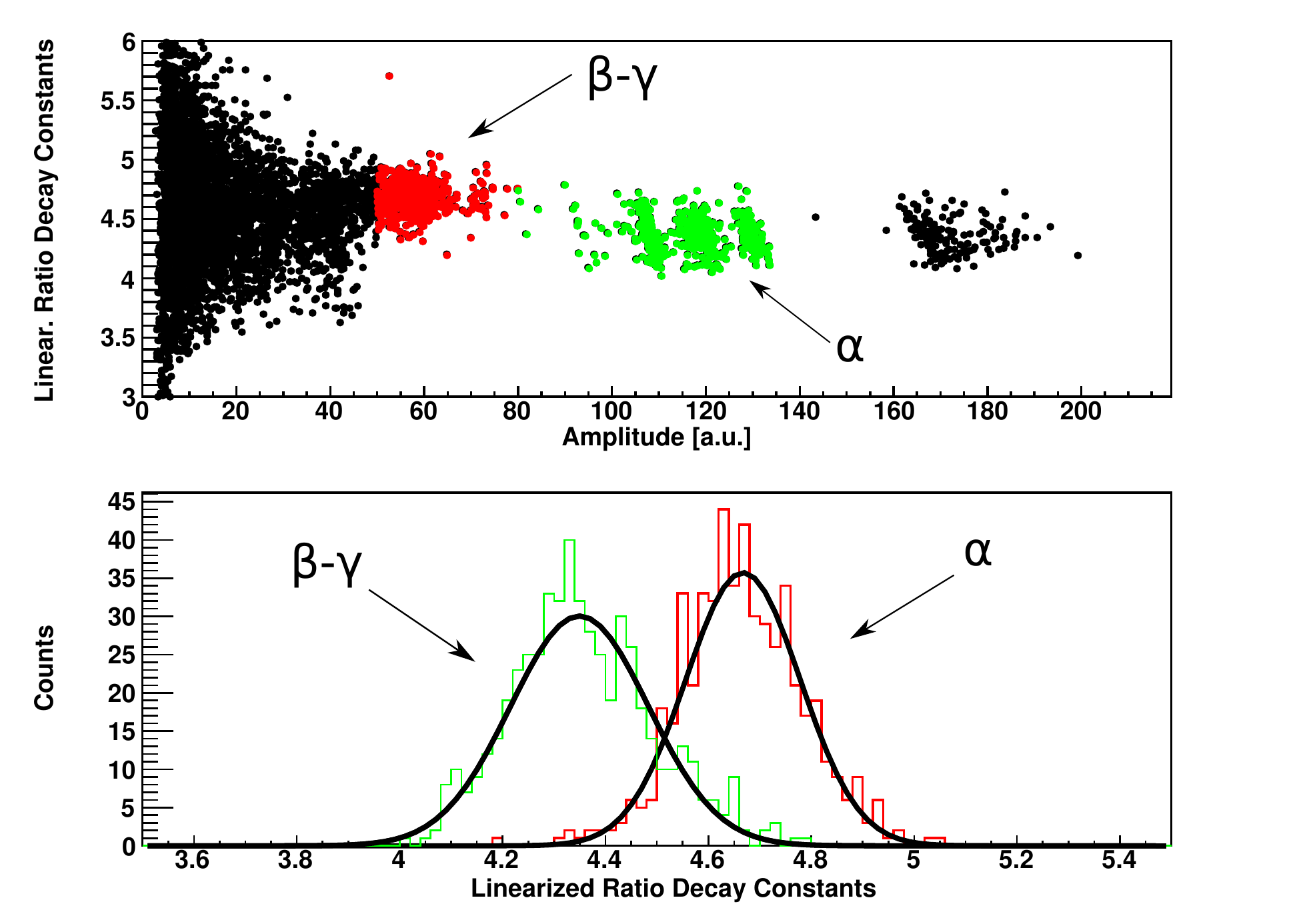}
\end{center}
\caption{MgMoO$_{4}$ crystal: scatter plot of RDC$^{lin}$ vs. Heat (top panel) and projection of 2-4 MeV events on the RDC$^{lin}$ axis (bottom panel). In green events in the 2-4 MeV region due to $\beta$/$\gamma$ particles and in red those due to $\alpha$'s.}
\label{fig:MgMoO4_RT}
\end{figure}

The crystal tested is 32x31x24 mm$^3$ with a weight of 89.1 g. The total live time of this measurements was $\sim$22 h. In this run it was not possible to face the crystal to a light detector because of the assembly structure so we can't use the Light vs. Heat scatter plot in order to tag $\alpha$ events. The performances of the bolometer were quite poor, most probably this was due to a problem with the gluing of the NTD thermistor: at the end of the measurement, when the crystal was back to room temperature, we discovered that under the thermistor the crystal showed a crack. This could explain why the signal to noise ratio was so bad (the energy resolution measured on $\alpha$ lines was \ca150 keV FWHM) and consequently also the resolution in the evaluation of the signal shape parameters was limited. Despite these problems, we were able to observe in the scatter plot of RDC$^{lin}$ vs. Heat (fig. \ref{fig:MgMoO4_RT}) clear difference between events due to the interaction of $\alpha$ particles and events due to $\beta$/$\gamma$ events. 
 
The discrimination between the two populations provided by the RDC parameter can be evaluated by means of a gaussian fit of the two peaks obtained by projecting selected events after  linearization. The resulting discrimination power D$_{RDC}$ is 1.8 sigma.

These preliminary results, even if limited because of the problems reported above, lead us to program new more detailed measurements to better study this promising crystal.

\subsection{ZnSe}

$^{82}Se$ is a \BB emitter with an isotopic abundance of 9.2\% and a Q-value of (2995.5 $\pm$ 2.7) keV. It has always been considered a good candidate for \BBz studies because of its high transition energy and the favorable nuclear factor of merit. For these reasons in the last years an R\&D work was carried out in which we have extensively studied the performances of ZnSe detectors in different conditions. For our studies we have used different ZnSe crystals. Characteristics of measurements done and obtained results are reported in detail in \cite{ZNSE}. Here we report the discrimination power on the RDC obtained with the Huge ZnSe crystal (h = 50mm, $\varnothing$ = 40mm, 337g). 

\begin{figure}[h]
\begin{center}
\includegraphics[ width=1.\linewidth]{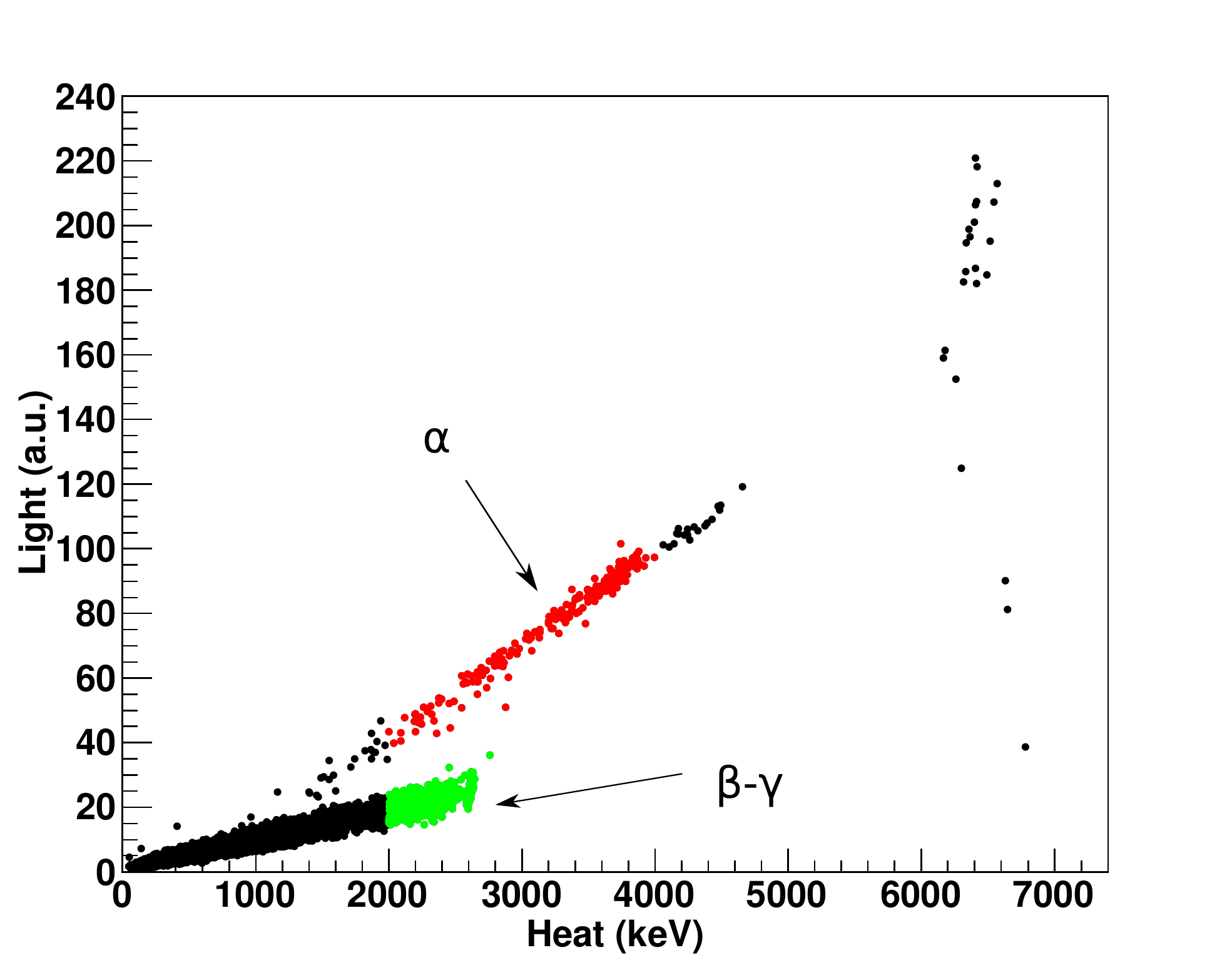}
\end{center}
\caption{ZnSe Crystal: scatter plot of Light vs. Heat.  In green events in the 2-4 MeV region due to $\beta$/$\gamma$ particles and in red those due to $\alpha$'s.}
\label{fig:znse}
\end{figure}

In Fig.~\ref{fig:znse} the Light vs. Heat scatter plot of a measurement of $\sim$70 h of live time is shown. In order to have a high number of $\alpha$ counts in the 2-3 MeV region also in these measurements a degraded $^{238}$U source was placed in front of the crystal. An explanation of the Quenching Factor larger than one (i.e. the $\alpha$ band lies above the $\beta$/$\gamma$ band) and other anomalies observed in this crystal can be found in \cite{ZNSE}.

\begin{figure}[h]
\begin{center}
\includegraphics[ width=1.\linewidth]{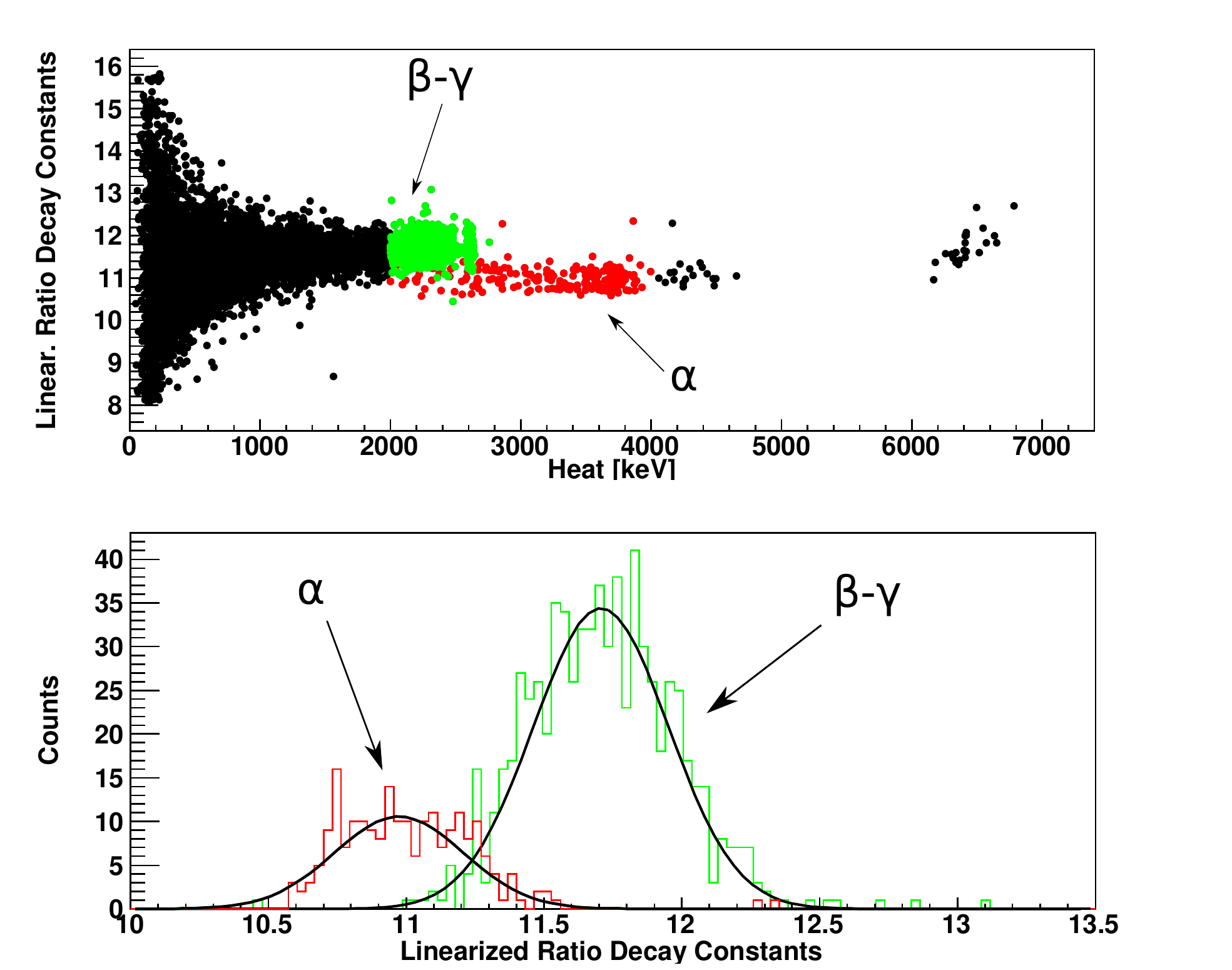}
\end{center}
\caption{ZnSe crystal: scatter plot of RDC$^{lin}$ vs. Heat (top panel) and projection of 2-4 MeV events on the RDC$^{lin}$ axis (bottom panel). In green events in the 2-4 MeV regions due to $\beta$/$\gamma$ particles and in red those due to $\alpha$'s.}
\label{fig:znse_RDC}
\end{figure}

Also in this case, the RDC$^{lin}$ vs. Heat (fig. \ref{fig:znse_RDC}) showed a difference between events due to the interaction of $\alpha$ particles and events due to $\beta$/$\gamma$ events. The discrimination power D$_{RDC}$ is 2.2 sigma.

\section{Conclusion}

The possibility to discriminate the nature of the particle interacting in a bolometric detector, simply on the base of the shape of the thermal pulse, is now definitely proved and opens new possibilities for the application of these devices in the field of rare events searches. In particular, the high rejection capability that could allow to completely rule out the $\alpha$ background in \BBz experiments was demonstrated. Unfortunately, based on the results obtained so far, the applicability of this technique to Dark Matter searches cannot yet be proved directly. This feature was observed in different scintillating crystals (CaMoO$_4$, ZnMoO$_4$, MgMoO$_4$ and ZnSe) and new tests are under preparation in order to investigate if a similar behavior can be observed also in other compounds.
A discrimination confidence level of $\sim$6.5 sigma was obtained both for CaMoO$_4$ and ZnMoO$_4$ crystals in the 2-4 MeV energy region. Discrimination confidence levels reached with MgMoO$_4$ (D$_{RDC}$=1.8$\sigma$) and ZnSe (D$_{RDC}$=2.2$\sigma$) are not so high but they indicate that even with these crystals the discrimination based on pulse shape analysis is possible.
New techniques aiming at improving the discrimination power of the pulse shape analysis are being studied.

\section{Acknowledgments}

This work was funded and developed under the Bolux experiment of INFN. Thanks are due to E. Tatananni,
A. Rotilio, A. Corsi and B. Romualdi for continuous and constructive help in the overall setup construction. Finally, we are especially grateful to Maurizio Perego for his invaluable help in the development and improvement of the Data Acquisition software.

\end{document}